\documentclass[11pt]{article}

\usepackage[T1]{fontenc}
\usepackage[margin=1in]{geometry}

\usepackage{graphicx}
\usepackage{subcaption}
\usepackage{physics}
\usepackage{qcircuit}
\usepackage{amsfonts}
\usepackage{bm}

\usepackage{amsmath,amssymb}
\usepackage{microtype}
\usepackage{authblk}
\usepackage{hyperref}

\hypersetup{
    colorlinks=true,
    linkcolor=blue,
    citecolor=blue,
    urlcolor=blue
}

\newcommand{\x}{{\mathbf{x}}}

\newcommand{\y}{{\mathbf{y}}}

\newcommand{\rv}{{\mathbf{r}}}

\newcommand{\B}{{\mathbf{B}}}
\newcommand{\Dx}{{\mathbf{D_x}}}
\newcommand{\Dy}{{\mathbf{D_y}}}

\newcommand{\SetX}{{\mathcal{X}}}
\newcommand{\SetY}{{\mathcal{Y}}}
\newcommand{\SetR}{{\mathcal{R}}}

\newcommand{\SetD}{{\mathcal{D}}}

\newcommand{\Rea}{{\mathbb{R}}}

\title{Quantum Minimal Learning Machine: A Fidelity-Based Approach to Error Mitigation}

\author{
Clemens Lindner$^{1}$\thanks{Email: \href{mailto:clindner@jyu.fi}{clindner@jyu.fi}},
Joonas Hämäläinen$^{1}$,
Matti Raasakka$^{2}$\\

$^{1}$University of Jyväskylä, Faculty of Information Technology, Finland\\
$^{2}$Aalto University, Department of Electronics and Nanoengineering, Finland
}

\date{}

\begin{document}

\maketitle

\begin{abstract}
    We introduce the concept of quantum minimal learning machine (QMLM), a supervised similarity-based learning algorithm. The algorithm is conceptually based on a classical machine learning model and adopted to work with quantum data. We will motivate the theory and run the model as an error mitigation method for various parameters.
\end{abstract}

\begingroup
\renewcommand{\thefootnote}{}
\footnotetext{Published version: \url{https://doi.org/10.1007/978-3-032-13852-1_30}}
\endgroup


\section{Introduction}
Quantum machine learning \cite{biamonte2017quantum} spans a wide range of possibilities: Leveraging fault tolerant quantum algorithms to accelerate core linear algebra routines \cite{harrow2009quantum}, adopting quantum counterparts of classical kernel methods \cite{schuld2021kernel}, and employing parametrized variational quantum algorithms (VQAs) optimized via a hybrid quantum-classical workflow \cite{jerbi2023quantum}. When run on actual quantum computing hardware these models suffer from hardware noise, therefore methods of error mitigation are becoming another important avenue for the mingling of machine learning with quantum computing \cite{liao2024machine}. In this paper we will introduce a novel similarity based quantum learning model and explore its use for error mitigation. We will take inspiration from a classical model known as minimal learning machine (MLM) \cite{de2015minimal} where a linear mapping between distances in the input and output spaces is used to learn the labels of some given data-set. Several quantum k-nearest neighbor approaches have been implemented \cite{zardini2024quantum,berti2024role}, yet most deal with the calculation of distances between classical data instead of quantum data \cite{kiani2022learning}. Since one of the hopes of quantum computers is that they will be able to simulate quantum systems and then provide us with synthetic data we aim to find a way of directly learning from that data. In section \ref{sec:1} we are first going to introduce the MLM model and then establish one possible way to use the same idea for quantum data. After we reformulated the model to work with quantum data we will use it in section \ref{sec:2} to demonstrate error mitigation of a given noisy quantum state. The main aim in this specific use-case will be to assume we are given a noisy quantum state from a real quantum computer where we don't know the precise noise model and then learn its ideal noise-free version. In our implementation we will simulate this process with depolarizing noise and a dataset of variational quantum circuits using Qiskit and QiskitAer. We will analyze the models performance with different circuit parameter settings (number of qubits, rotational parameter ranges, Ansatz layers, noise levels) and provide the corresponding code.

\section{Conceptual framework and preliminaries} \label{sec:1}

\subsection{MLM}

Minimal learning machine (MLM) is a supervised learning method useful for learning the relationship between a set of input and output points. An exemplary use-case is multi label classification (MLC) \cite{hamalainen_minimal_2025} where the output points correspond to labels of the input points. In the following we will work with that example. \\
\\
We are given a dataset $\SetD = \{ (\x_1,\y_1), \dots, (\x_N,\y_N)\}$, where $\x_i \in \Rea^M$ corresponds to an input vector and $\y_i \in \{0,1\}^{L}$ to the multi-hot encoded output label vector for $i = 1, \dots, N$. In $\y_i$, value $1$ at index $j$ represents that class-label $j$ is associated with instance $i$. As an example, $\y_1=(1,1,0,0)$ could correspond to the vectorized picture data $\x_1$ including class-label 1 and 2 but not 3 and 4. For a MLC dataset, $\sum_{j = 1}^{L} \y_i(j) > 1$ at least for one instance.\\
\\
The approach goes as follows: We calculate the distances between the input space points and some set of reference points, this is done accordingly in the output space. We then store them in distance matrices $\Dx$ and $\Dy$ which are correlated via a parametrized matrix $\B$, i.e. $\Dx \B=\Dy$. After having learned $\B$ with a set of training points we can use it to label new data points.\\
\\
In MLM, the core subroutine is to solve the model's coefficient matrix $\B$ via ordinary least squares $\B = (\Dx^T\Dx)^{-1}\Dx^T\Dy$ for feature space and output space distance matrices, $\Dx \in \Rea^{N \times K}$ and $\Dy \in \Rea^{N \times N}$, where $\Dx_{(i,j)} = \|\x_i-\rv_j\|_2$, $\rv_j \in \SetR = \{\rv_k \}_{k=1}^K \subseteq \SetX$ and $\Dy_{(i,j)} = \|\y_i-\y_j\|_2$. The quality of the linear mapping $\B$ is affected by how feature space reference points $\SetR$ are selected. If $\Dx$ is a full rank matrix for $K=N$, we can use $\B = \Dx^{-1}\Dy$. In the prediction phase of MLM, for a given instance $\x$ distances are computed to $\SetR$ and then $\B$ is used to estimate output space distances to all training labels. An actual prediction for a given task is then extracted by post-processing of the distance estimates. The simplest approach for this is to identify the smallest predicted distance and its corresponding label vector $\hat y \in \SetY$ \cite{hamalainen_minimal_2025}. Note that by selecting $\SetR = \SetX$ ($K = N$), MLM is deterministic.

\subsection{QMLM}

In a 'quantized' version of MLM we can use the same model to work with quantum data. For illustrative purposes we keep with the multi label approach and introduce the space $\mathcal{H}_\SetX$, a Hilbert space in which our input data lives. More specifically the input data we wish to prescribe multiple labels to is a set of pure quantum states $\ket{\Psi}_i \in \mathcal{H}_\SetX$ with $i \in \{1,2, \dots, N\}$ or their corresponding density matrices $\rho_i=\ketbra{\Psi_i}$. As soon as we deal with quantum data we have to use a different distance measure to quantify data similarity. An intuitive way is to introduce the fidelity between two quantum input states as our distance measure, i.e.
\begin{equation} \label{fide}
    F(\ket{\Psi}_i, \ket{\Psi}_j) = |\braket{\Psi_i}{\Psi_j}|^2, \quad \text{where } \ket{\Psi}_i, \ket{\Psi}_j \in \mathcal{H}_\SetX.
\end{equation}
We have $F(\ket{\Psi}_i, \ket{\Psi}_j)=F(\ket{\Psi}_j, \ket{\Psi}_i)$ and $0 \leq F(\ket{\Psi}_i, \ket{\Psi}_j) \leq 1$ where a value closer to one means the states are more similar. We choose $\SetR = \SetX$ and therefore define the matrix $\SetD_{\mathcal{H}_\SetX}$ to hold the fidelity values between all pairs of quantum states in the input space:

\begin{equation} \label{DX}
D_{\mathcal{H}_\SetX} = \begin{bmatrix}
F_{11} & F_{12} & \cdots & F_{1N} \\
F_{21} & F_{22} & \cdots & F_{2N} \\
\vdots & \vdots & \ddots & \vdots \\
F_{N1} & F_{N2} & \cdots & F_{NN} \\
\end{bmatrix},
\end{equation}
where $F_{ij}=F(\ket{\Psi}_i, \ket{\Psi}_j)$ and $i, j \in \{1, 2, \dots, N\}$. In practice, these fidelity values can be determined by implementing a swap test.\\
\\
For the multi label case we could naively assume an encoding of the labels through the computational basis states, i.e. $\ket{\Phi_i}=\ket{01101} \in \mathcal{H}_\SetY$. This approach however would lead to fidelity values of either 0 or 1 and therefore our distance matrix in the label space wouldn't hold much information. We also have to find a way where a large Hamming distance between two label vectors, i.e. the number of bit positions where they differ $d_H(\y_i,\y_j)$, corresponds to a low fidelity between the two corresponding quantum states. This can be achieved by using two overlapping states in our representation. We encode the label vector $\y_i$ in the quantum state $\ket{\Phi_i}$ where $i\in \{1, 2, \dots, N\}$. If we denote the $k$-th value in a bit string of length $L$ corresponding to the label $\y_i$ as $b_{ik}$ and it's corresponding quantum state as $\ket{\phi_{ik}}$ we can write

\begin{equation}
    \ket{\Phi_i}=\bigotimes_{k=1}^{L} \ket{\phi_{ik}} = \bigotimes_{k=1}^{L} 
\begin{cases}
\ket{+}, & \text{if } b_{ik} = 1 \\
\ket{0}, & \text{if } b_{ik} = 0,
\end{cases}
\end{equation}
\\
where $\ket{+}=\frac{1}{\sqrt{2}}(\ket{0}+\ket{1})$. The bit string 01101 would hence be represented as $\ket{\Phi_i}=\ket{0++0+}$. Since $\braket{0}{0}=\braket{+}{+}=1$ and $\braket{0}{+}=\braket{+}{0}=\frac{1}{\sqrt{2}}$ we have $\braket{\Phi_i}{\Phi_j}=(\frac{1}{\sqrt{2}})^{d_H(\y_i,\y_j)}$ and therefore

\begin{equation}
    F(\ket{\Phi_i}, \ket{\Phi_j})=|\braket{\Phi_i}{\Phi_j}|^2=\left( \dfrac{1}{2} \right)^{d_H(\y_i, \y_j)}.
\end{equation}
This way a large Hamming distance leads to a low fidelity value, as intended. We denote the distance matrix in the output space as $D_{\mathcal{H}_\SetY}$ with entries 
\( F'_{ij} = F(\ket{\Phi_i}, \ket{\Phi_j}) \) where \( i, j \in \{1, 2, \dots, N\} \). The rest of the algorithm can follow analogously to the classical MLM. We assume a linear mapping between the distance spaces, i.e. $D_{\mathcal{H}_\SetX} \mathcal{B}=D_{\mathcal{H}_\SetY}$ and obtain $\mathcal{B}$ by taking the pseudoinverse of $D_{\mathcal{H}_\SetX}$. We then calculate the fidelities for a new test data state $\ket{\Psi_{t}}$ to all other input states, apply the mapping $\mathcal{B}$ and gain a vector encoding all mapped similarities in the label space. Since this vector encodes fidelities we now have to take the maximum value which corresponds to the highest similarity. The index of that value will give us the label corresponding to $\ket{\Psi_t}$, i.e. $\ket{\Phi_t}$, which after a quick remapping ($\ket{+}\mapsto 1, \ket{0} \mapsto 0$) will yield the labels bit string.\\
\\
While this specific QMLM example could be used for multi label classification of quantum states it is primarily useful if we want to have the labels stored on a quantum computer and could still be improved upon by e.g. choosing a different label encoding. In the next section we are going to discuss a more near term use-case for QMLM, that being the use for error mitigation of noisy quantum states. Looking further ahead to fault-tolerant devices, QMLM may also prove valuable in any context where a systematic relationship can be established between the similarity structure of quantum states and that of a corresponding output space.

\section{QMLM for error mitigation} \label{sec:2}

In this section the aim of our QMLM model will be to learn an ideal quantum state when provided with its noisy version. The assumption is that we are given a set of noisy quantum states which could have been obtained from running quantum circuits that suffer from some hardware-specific noise. Our dataset consists of noisy states as inputs, i.e. $\{\rho_{i,{\text{noisy}}}\}$ and the corresponding ideal pure states $\{\ket{\Psi_i}\}$ as outputs. The fidelity in the input space would therefore take the general form
\begin{equation} \label{Frho}
    F(\rho_i, \rho_j) = \left( \text{Tr} \sqrt{\sqrt{\rho_i} \rho_j \sqrt{\rho_i}} \right)^2,
\end{equation}
while the fidelity in the output space reduces to the form we have in equation \ref{fide}. The assumption is now that since there always exists a completely positive trace preserving map $\mathcal{E}$, s.t. $\mathcal{E}(\ketbra{\Psi_{\text{ideal}}})=\rho_{\text{noisy}}$, we can learn a useful relationship between the similarity matrices in the input and output space that could provide us with information about the specific hardware noise we encounter. In our implementation of the model we focused on depolarizing noise, i.e.
\begin{equation}
\mathcal{E}_{\text{depol}}(\rho_i) = (1 - p) \rho_i + \frac{p}{d} I
\end{equation}
where \( \rho_i \) is the initial quantum state, \( p \) is the depolarizing probability with \( 0 \leq p \leq 1 \), \( d \) is the dimension of the Hilbert space (for an \( n \)-qubit system, \( d = 2^n \)) and \( I \) is the \( d \times d \) identity matrix. Let \( \rho_1' \) and \( \rho_2' \) be the density matrices obtained after applying the depolarizing noise channel \( \mathcal{E}_{\text{depol}} \) to the initial pure states \( \rho_1 \) and \( \rho_2 \) with depolarizing parameters \( \lambda_1 \) and \( \lambda_2 \), respectively:

\begin{equation}
\rho_i' = (1 - \lambda_i)\,\rho_i + \frac{\lambda_i}{d}\,I,
\quad i = 1,2\,.
\end{equation}
\\
Reordering for $\rho_1$ and $\rho_2$ and taking the fidelity formula for pure states $F(\rho_1, \rho_2) = \text{Tr}(\rho_1\rho_2)$, we can relate the fidelity of our pure states with the trace over the resulting mixed ones:

\begin{equation}
\text{Tr}(\rho_1'\rho_2') = \alpha F(\rho_1, \rho_2) + \frac{1 - \alpha}{d},
\end{equation}
where $\alpha = (1 - \lambda_1)(1 - \lambda_2)$. In theory, this would therefore motivate a mapping between fidelities for ideal states and the traces of their noisy counterparts which shows similarities to the approach found in \cite{czarnik2021error}, where a linear model is motivated for ideal and noisy expectation values. The benefit of our model (which corresponds to each ideal fidelity being a linear combination of noisy fidelities) is that it could model the underlying noise more generally. This could be useful since we are in reality never limited to just one specific noise model. In our setting, once the ideal version of a quantum state is obtained, the noise-free expectation value of an operator $O$ can be evaluated via $\langle O \rangle_{\text{ideal}} = \mathrm{Tr}\!\left( O \rho_{\text{ideal}}\right)$.\\
\\
Although our QMLM model has the benefit of broad applications, it does suffer from the same problem as quantum kernel methods, that being the problem of exponential concentration \cite{thanasilp2024exponential}. This is obvious since the fidelity matrix in equation \ref{DX} represents a Gram matrix whose off diagonal values go to zero if the dimension of the Hilbert space grows too large. In our implementation, we try to negate this effect by limiting the region of the accessible Hilbert space through small parameter changes in a specified variational quantum circuit. For real applications this could correspond to the knowledge that a set of given quantum data is limited to a certain region of Hilbert space or exhibits some inherent symmetry akin to approaches in geometrical quantum machine learning \cite{ragone2022representation}.

\subsection{Qiskit Implementation}

\begin{figure}[!htb]
  \centering

  \begin{subfigure}[t]{0.495\columnwidth}
    \includegraphics[width=\linewidth]{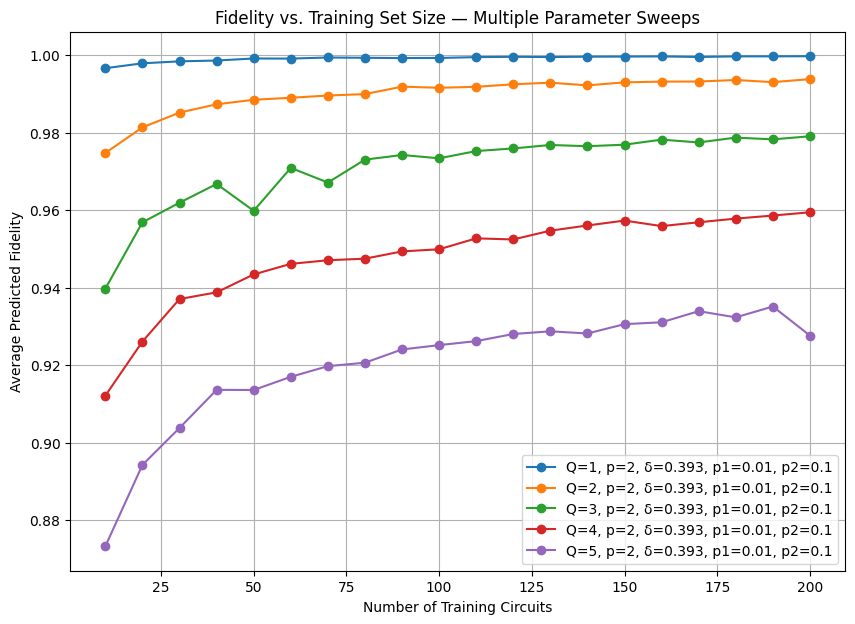}
    \caption{Varying qubit numbers}
    \label{fig:qubits}
  \end{subfigure}\hfill
  \begin{subfigure}[t]{0.495\columnwidth}
    \includegraphics[width=\linewidth]{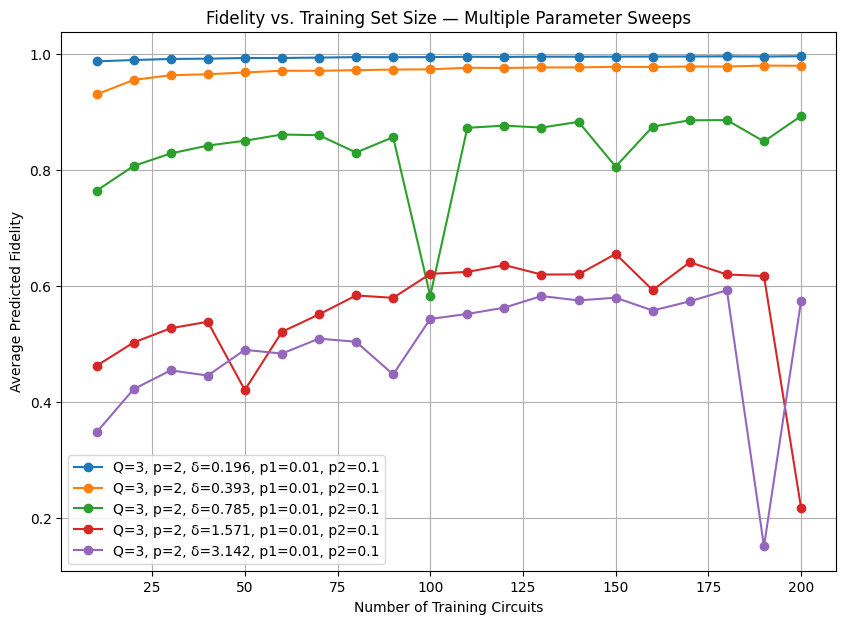}
    \caption{Varying \(\theta\) ranges}
    \label{fig:deltas}
  \end{subfigure}

  \vspace{1em}

  \begin{subfigure}[t]{0.495\columnwidth}
    \includegraphics[width=\linewidth]{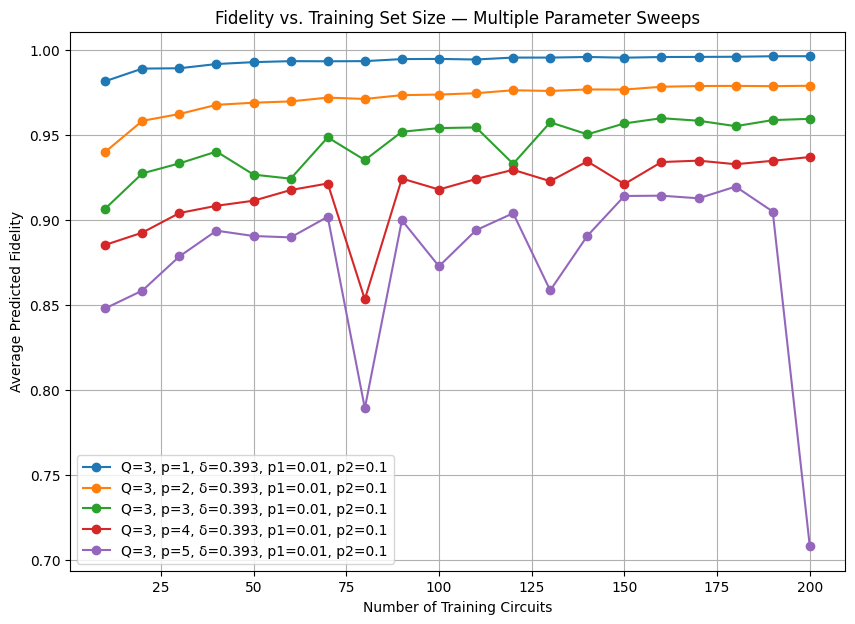}
    \caption{Varying Ansatz layers}
    \label{fig:p}
  \end{subfigure}\hfill
  \begin{subfigure}[t]{0.495\columnwidth}
    \includegraphics[width=\linewidth]{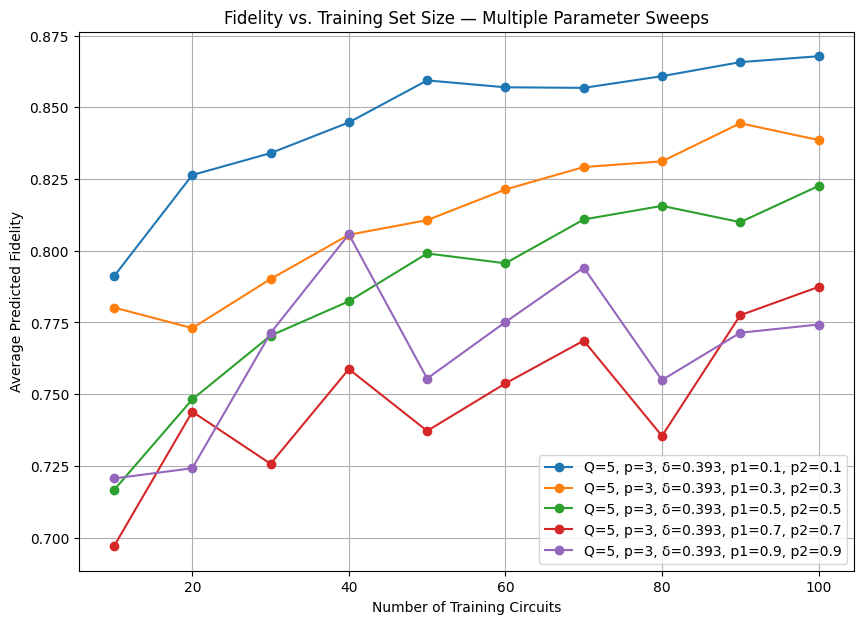}
    \caption{Varying noise levels}
    \label{fig:noise}
  \end{subfigure}

  \caption{Average predicted fidelity vs.\ dataset size under different experimental parameters.}
  \label{fig:all-results}
\end{figure}

We implement the depolarizing noise model with QiskitAer and use the same noise parameters for each training data set. Our output space dataset consists of $N$ variational quantum circuits with an Ansatz corresponding to the following form:

\[
\Qcircuit @C=1em @R=1em {
\lstick{\ket{q_0}} & \gate{R_x(\theta_1)} & \gate{R_z(\theta_2)} & \ctrl{1} & \qw \\
\lstick{\ket{q_1}} & \gate{R_x(\theta_3)} & \gate{R_z(\theta_4)} & \targ    & \ctrl{1} \\
\lstick{\ket{q_2}} & \gate{R_x(\theta_5)} & \gate{R_z(\theta_6)} & \qw      & \targ \\
}
\]
\\
Because of the already mentioned vanishing fidelity problems, we will adjust the range of the randomly drawn $\theta$ from $\theta \in[-\pi,+\pi]$ to smaller intervals given by some $\delta$, e.g. $\delta=\frac{\pi}{2}$ s.t. $\theta \in [-\delta,+\delta]$. The amount of repeated Ansatz layers can be specified by a parameter $p$, the depicted circuit here corresponds to $p=1$. More specifically our output space dataset is then given by the statevectors we gain from such circuits using an ideal simulator. For each VQA circuit of the output space dataset we apply the depolarizing noise model which applies some error to each 1 and 2 qubit gate in the circuit depending on the chosen parameters $p_1$ and $p_2$. For each circuit we then gain the mixed density matrix as an end result of that noisy simulation and set the input space to be the collection of those.\\
\\
After creating our dataset of ideal statevectors (the set $\SetY$) and their noisy counterparts (the set $\SetX$) we calculate their corresponding fidelity matrices $D_{\mathcal{H}_\SetY}$ and $D_{\mathcal{H}_\SetX}$. Through a standard pseudoinverse of $D_{\mathcal{H}_\SetX}$ we gain $\mathcal{B}$. Now a new data point is created by randomly drawing a new set of $\theta_i$ angles and applying the noise model to the corresponding circuit, meaning the new data point is some noisy state but by the nature of our process we have access to the ideal one. In reality such a new noisy state would be obtained from an actual quantum computation. After the QMLM procedure, we gain the quantum state of our set $\SetY$ which exhibits the highest similarity to the ideal version of the new data point. This suggests that using more training samples improves performance, where performance is quantified by the fidelity between the predicted and the true ideal state. Because our data is generated randomly, we evaluate performance across many new test circuits and report it as the “average predicted fidelity”.\\
\\
In Fig. \ref{fig:qubits} we provide a result for that fidelity in comparison with the number of training dataset points (i.e. the number of VQA circuits) for $Q\in [1,2,3,4,5]$ qubits. For each visible point in all graphs we have averaged over 400 predictions. We limited the rotation range to $\theta \in [-\frac{\pi}{8},+\frac{\pi}{8}]$ and have increased the noise to a level above known real hardware noise (approximately $p_1=0.001$ and $p_2=0.01$) to make its effect more prevalent. We can see that a higher dimensional Hilbert space leads to worse performance since we would need more points in order to approximately cover the whole space. Also while there is an initial increase in performance we can also see a leveling effect as soon as the reachable space is sufficiently covered by the training set data.

\noindent 
In Fig. \ref{fig:deltas} we compare different $\delta \in [ \pi, \frac{\pi}{2}, \frac{\pi}{4}, \frac{\pi}{8}, \frac{\pi}{16}]$ values and focus on three qubits. Despite the averaging we can see some statistical fluctuations which should be smoothed if we increase the number of predictions we average over while increasing model runtime. We observe that when the accessible Hilbert‐space region grows, performance degrades, reflecting that both space dimension and VQA circuit expressibility impact results. Since more Ansatz layers also correspond to higher expressibility, the same effect should be a cause of worse performance as seen in Fig. \ref{fig:p} where we compare different Ansatz layer parameters $p$. The worse performance for higher $p$ values can also in general be traced to the fact that any gate errors now have more chances to accumulate.\\
\\
At last we will also compare different noise levels in Fig. \ref{fig:noise}. Regarding the noise levels, our model runs into two singular cases: one where the noise is zero and $\mathcal{B}$ therefore becomes the identity matrix and the other one where the noise is maximal, leading to all input states being the maximally mixed one. In the first case the model basically acts as a lookup table, i.e. it checks if the new state is in the training data whereas the second case corresponds to $D_{\mathcal{H}_\SetX}$ consisting of only 1's in its entries which leads to the output similarity vector consisting of the same value in each entry hence providing us some trivial average performance depending on $D_{\mathcal{H}_\SetY}$. We plot the results for 5 qubits and note that the noise levels we chose are unrealistic, however increasing them still shows a general decrease in the models performance. This probably corresponds to the fact that the more noise we introduce the more similar the states become, leaving us with less information to learn from. We also observed in general larger fluctuations when increasing noise levels.\\
\\
While we laid the groundwork for the QMLM model, further improvements can definitely still be done. Examples would be to change the post processing from a simple nearest neighbor approach to some weighted average over the closest points or to test if a different mapping between the distance matrices could lead to a better performance. Furthermore it would be interesting to run the model on actual quantum hardware and see if it sufficiently captures the more general noise models of real case scenarios. This however would lead to other problems, e.g. we can't just run a swap test on real hardware to measure the fidelity of noisy states since the swap test quantum gates themselves would be affected by the noise. We  therefore either have to mitigate such effects or access some other methods like quantum state tomography.

\section{Conclusion}

In this paper, we presented an approach towards a 'quantized' MLM model. This was done by assuming the availability of a given set of quantum states and using the fidelity between them as a similarity measure. We specified use cases and ran the model for error mitigation. Our results show a good performance as long as we limit the accessible range of our created quantum states, more work is however still needed to test if the model is usable for higher dimensional data. QMLM could be a versatile approach to learning features of quantum states and a possible future avenue for learning immediately from the quantum data created on a fault-tolerant quantum computer.

\section{Code availability}

Our code was run in Jupyter notebooks using Qiskit and QiskitAer for the quantum simulations. The full implementation is provided in the following GitHub: \url{https://github.com/clemensLin/QMLM}.

\section*{Acknowledgements}

We acknowledge the financial support of the Finnish Ministry of Education and Culture through the Quantum Doctoral Education Pilot Program (QDOC VN/3137/2024-OKM-4) and the Research Council of Finland through the Finnish Quantum Flagship project (JYU 359240, University of Jyväskylä).

\section*{Competing interests}

The authors have no competing interests to declare that are relevant to the content of this article.

\bibliographystyle{ieeetr}
\bibliography{refs}

\end{document}